\documentclass[3p]{elsarticle}

\usepackage{amssymb}
\usepackage{amsmath}
\usepackage{upgreek}
\usepackage{booktabs}

\usepackage{orcidlink}

\usepackage{color}

\usepackage{hyperref}

\usepackage{graphicx} 

\biboptions{sort&compress}

\begin{document}

\begin{frontmatter}

\title{Unified machine-learning framework for property prediction and time-evolution simulation of strained alloy microstructure}

    \author[1]{Andrea Fantasia \orcidlink{0009-0009-1169-5083}}
\author[1,2]{Daniele Lanzoni \orcidlink{0000-0002-1557-6411}}
    \author[1]{Niccolò Di Eugenio \orcidlink{0009-0001-0682-855X} \fnref{fn1}}
\author[1]{Angelo Monteleone}
\author[1]{Roberto Bergamaschini \orcidlink{0000-0002-3686-2273}}
\author[1]{Francesco Montalenti \orcidlink{0000-0001-7854-8269}}
\ead{francesco.montalenti@unimib.it}

\affiliation[1]{organization={Department of Materials Science, University of Milano-Bicocca}, 
            addressline={Via R. Cozzi 55}, 
            postcode={20125}, 
            state={Milano},
            country={Italy}}

\affiliation[2]{organization={Department of Physics, University of Genova}, 
            addressline={Via Dodecaneso 33}, 
            postcode={16146}, 
            state={Genova},
            country={Italy}}

\fntext[fn1]{Present address: INFN and Dept. of Applied Science and Technology, Polytechnic University of Turin, 10129, Turin, Italy}

\begin{abstract}
We introduce a unified machine-learning framework designed to conveniently tackle the temporal evolution of alloy microstructures under the influence of an elastic field. This approach allows for the simultaneous extraction of elastic parameters from a short trajectory and for the prediction of further microstructure evolution under their influence. This is demonstrated by focusing on spinodal decomposition in the presence of a lattice mismatch $\eta$, and by carrying out an extensive comparison between the ground-truth evolution supplied by phase field simulations and the predictions of suitable convolutional recurrent neural network architectures. The two tasks may then be performed subsequently into a ``cascade'' framework. Under a wide spectrum of misfit conditions, the here-presented cascade model accurately predicts $\eta$ and the full corresponding microstructure evolution, also when approaching critical conditions for spinodal decomposition. Scalability to larger computational domain sizes and mild extrapolation errors in time (for time sequences five times longer than the sampled ones during training) are demonstrated. The proposed framework is general and can be applied beyond the specific, prototypical system considered here as an example. Intriguingly, experimental videos could be used to infer unknown external parameters, prior to simulating further temporal evolution. 
\end{abstract}

\end{frontmatter}

\section{Introduction} \label{sec::intro}
Modelling microstructural features and their dynamics is key for materials science and engineering due to their fundamental impact on functional and mechanical properties~\cite{kok2018anisotropy, li2020mechanical}. A full characterization of these aspects is generally demanding from an experimental point of view, especially when involving time-dependent properties, thus making computational approaches particularly useful. Indeed, a plethora of simulation approaches have been developed for decades on this topic~\cite{kim2000computation, elder2007phase, moelans2008introduction}, but the problem remains challenging for several reasons. First, high-accuracy methods usually bring with them also high computational costs. Second, the connection between models and experiments often requires difficult, indirect comparisons. Finally, understanding microstructure properties is inherently a multi-scale and multi-physics problem, thus involving a large number of parameters and conditions hard to control and identify. 

In recent years, ML and, most prominently, Neural Networks (NN), have revolutionized materials science, opening new ways for materials discovery \cite{lyngby_data-driven_2022}, design \cite{ zhao_physics_2023, chenebuah_deep_2024, karpovich_deep_2024}, and property prediction \cite{ward_general-purpose_2016, dunn_benchmarking_2020, de_breuck_materials_2021}. ML-driven interatomic potentials changed the rules of atomistic modelling, delivering accuracy on par with first-principle approaches at a fraction of the computational costs \cite{bartok_gaussian_2010, behler_atom-centered_2011, fantasia_development_2024}. Recent studies have also highlighted the potential for optimizing material fabrication processes through real-time feedback control \cite{shen_-situ_2024, shen_autonomous_2024}, showcasing the synergy between ML and experimental techniques.

At the mesoscale, NNs have demonstrated exceptional power for accelerating~\cite{strayerADDMANL2022,lanzoniAPLML2024}, or fully surrogating~\cite{montes_de_oca_zapiain_accelerating_2021, hu_accelerating_2022, wu_emulating_2023, ahmad_integrated_2024, lanzoni_extreme_2024,ren_numerical_2022, ren_phase-field_2022, lee_recent_2023, wang_modeling_2024,choi_accelerating_2024, alhadaNPJCM2024} conventional simulation approaches, eventually predicting time-evolutions without resorting to the explicit solution of the underlying partial differential equations. The typical task of such works has been to train a NN model capable of recognizing spatial- and time-correlations within suitably processed time-series of data, either pixelated images of the relevant fields or some latent-space representation of them, and then predict evolution frames starting from an arbitrary configuration. In the simplest approaches, the NN model is trained to learn the solution for an assigned parameter set and then it is used to predict the corresponding evolution as a function of the different initial configuration, offering substantial speed-ups in the solution compared to the explicit numerical scheme, eventually giving access to large-scale domains or long-times \cite{lanzoni_extreme_2024, yang_self-supervised_2021, fan2024accelerate} that exceed the training dataset characteristics. More advanced studies extended the training over more varied datasets, comprising cases generated by the same dynamics but sampling a "broad" range of values of some constitutive parameters, responsible for qualitative variations in the evolution. In such a case, the NN is expected to implicitly identify the parameters by processing a short initial evolution sequence and then continuing it at later stages \cite{hu_accelerating_2022, yang_self-supervised_2021}. It is worth noticing, however, that in this approach, the NN does not provide any estimation of the actual value of the conditioning parameters as they remain encoded in the latent representation of the input data. Last, a few recent works \cite{guptaARXIV2022,oommen_rethinking_2023} implemented explicit conditioning by external parameters (or boundary conditions \cite{alhadaNPJCM2024}) within different NN architectures so as to control the predictive process by supplying them in input along with the initial configuration.

In this work we propose a unifying approach based on a Convolutional Recurrent Neural Network (CRNN) architecture \cite{lanzoni_morphological_2022, lanzoni_extreme_2024} with an explicit treatment of a physical parameter controlling microstructural evolution, capable of: (1) inferring the unknown value of such parameter by processing a, possibly short, time-series; (2) accurately predicting the time evolution of an arbitrary initial configuration, conditioned by the known, from (1) or a priori, parameter. Compared to the previously cited literature, the fact of preserving the identity of the physical parameter, rather than letting the network infer it implicitly, makes the model much easier to interpret. Moreover, this opens the way for a cascade usage scenario starting from the estimation of the unknown parameters by the analysis of evolution images, eventually acquired from (scarcely available) experiments, and then using the CRNN for extensive simulations based on those. Importantly, the trained model allows for generalizations on both temporal and spatial scales, thus giving the possibility of scouting cases beyond the training conditions.

As a widely-investigated prototypical case~\cite{oommen_rethinking_2023, lanzoni_extreme_2024, yang_self-supervised_2021}, we here consider the phenomenon of spinodal decomposition, a second-order phase transition by which a homogeneous mixture separates into two distinct phases after cooling below a critical temperature~\cite{cahnACTAMETAL1961,langer_theory_1973, chenBOOK1994,fratzl_modeling_1999,kwon_coarsening_2007, andrews_effect_2020}.
This mechanism plays a critical role in various systems of metallurgical interest, including Al-Zn alloys \cite{rundman_early_1967, xu_stabilizing_2024}, Ni-based superalloys \cite{collins_spinodal_2020}, and certain oxides \cite{kim_spinodal_2002, ban_spinodal_2023}. Recent interest in spinodal behaviour has also been reignited by the design and study of spinodoid metamaterials \cite{kumar_inverse-designed_2020, zhengCMAME2021}. In crystalline materials, the two emerging phases typically differ in molar volume, leading to the development of elastic strain during the transformation. This strain can give rise to characteristic microstructural patterns, particularly in systems with anisotropic or spatially inhomogeneous elastic properties \cite{nishimoriPRB1990, fratzl_modeling_1999, zhuMSME2001}. Representative examples of microstructural evolutions obtained from our simulations for a binary mixture with cubic elastic constants \cite{cahnACTAMETAL1962}, are shown for different mean composition and misfit $\eta$. Given the evident morphological changes from rounded to elongated domains for increasing misfit, $\eta$ is selected as an explicit parameter in our CRNN model, to be predicted from a time series or introduced as input to predict the correct evolution. The model also implicitly learns all other dependencies from composition and geometry through the NN training on a heterogeneous dataset.

\begin{figure}[t!]
 \centering
   \includegraphics[width=0.9\linewidth]{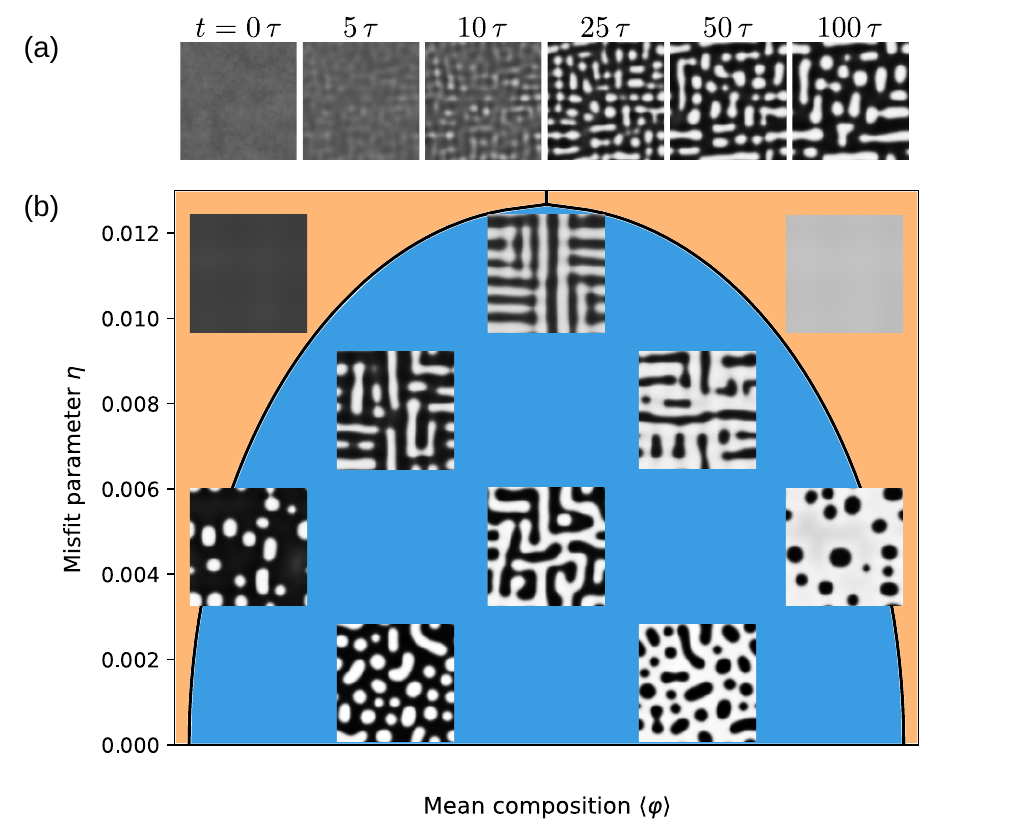}
    \caption{(a) Example of a phase-field evolution from the dataset, starting from a Perlin noise initial condition and corresponding to a misfit strain of $\eta \approx 0.48\%$. (b) Examples of the different morphologies of the spinodal decomposition of a binary alloy with cubic elastic anisotropy, as a function of mean composition $\left\langle \varphi \right\rangle$ and misfit strain $\eta$. The reported images correspond to representative evolution stages obtained by the numerical integration of the CH equation~\eqref{eq::spectral} starting from the homogeneous phase slightly perturbed by Perlin noise around the average composition $\left\langle \varphi \right\rangle$. The theoretical spinodal region (blue area) is reported as a reference in the background.} \label{fig::data}
\end{figure}

The paper is structured as follows. In Sect.~\ref{sec::methods} we first define the physical model for the microstructure, based on a Phase-Field (PF) description (\ref{sec::methods::pf_mod}), detail the dataset preparation process (\ref{sec::methods::data}), and then outline the CRNN architecture, training methodology, and evaluation framework (\ref{sec::methods::nn_archi}). The results are then discussed in Sect.~\ref{sec::results}. In particular, we first analyse the NN performances separately for the task of parameter estimation (\ref{sec::results::para}) and for predicting the time-evolution using it (\ref{sec::results::evo}). Then, we combine the two stages into a cascade approach (\ref{sec::results::cascade}). In all cases, we evaluate the reliability of the NN with respect to the ground truth numerical simulations as well as the generalization performance of the model. Finally, we draw conclusions and discuss the potential uses for the proposed method.

\section{Methods} \label{sec::methods}

\subsection{PF model of coherently strained spinodal decomposition} \label{sec::methods::pf_mod}
The standard way for modelling spinodal decomposition dynamics is by the phase field approach \cite{provatas_phasefield_2010}. For isothermal conditions, the system configuration is specified by an order parameter $\varphi$, tracing the local composition of the binary mixture, and by the total strain tensor field $\boldsymbol{\varepsilon}(x)$. Linear elasticity is considered throughout the paper. The system free energy $F$ is then expressed by the Ginzburg-Landau functional \cite{garcke2003}:
\begin{equation}\label{eq::FreeEnergy}
    F[\varphi, \boldsymbol{\varepsilon}] = \int_\mathcal{D} \left ( \frac{\epsilon}{2} | \vec\nabla \varphi|^2 + w(\varphi) +  \rho(\varphi,\boldsymbol{\varepsilon}) \right ) \, d\vec{x} \, ,
\end{equation}
where $\mathcal{D}$ is the physical domain, $\epsilon$ sets the width of the interface between the two phases. The $|\vec{\nabla} \varphi|^2$ term accounts for the energy cost of such interfaces, while the remaining ``bulk'' contribution divides into a double-well potential
\begin{equation}
    w(\varphi) = \frac{18}{\epsilon} \varphi^2 (1-\varphi)^2 \, ,
\end{equation}
with minima at $\varphi = 0$ and $\varphi = 1$, corresponding to the equilibrium compositions of the two separated phases in the absence of strain \footnote{A fixed temperature is here assumed so that $\varphi$ can be considered a rescaling of the full composition range $c$ to the $[c_1,c_2]$ one of the two phases at that temperature, i.e. $\varphi=\frac{c-c_1}{c_2-c_1}$}, and an elastic energy density $\rho (\varphi,\boldsymbol{\varepsilon})$:
\begin{equation}\label{eq::elastic}
    \rho(\varphi,\boldsymbol{\varepsilon}) = \frac{1}{2} C_{ijkl} (\varepsilon_{kl}-\varepsilon_{kl}^*(\varphi))(\varepsilon_{ij}-\varepsilon_{ij}^*(\varphi)) \, ,
\end{equation}
where $C_{ijkl}$ is the fourth-order tensor of elastic constants and $\boldsymbol{\varepsilon^*}$ is the eigenstrain tensor, i.e. the zero-stress strain \cite{mura1987}. Einstein notation, implying summation of repeated indices, is adopted. The $\rho$ term accounts for the different lattice parameters of the pure materials composing the alloy. Due to this additional energy cost, the bulk energy minima tend to move inward of the [0,1] range.

In the present study, we consider (elastically) homogeneous materials, i.e. with $C_{ijkl}$ independent of $\varphi$, and purely dilational eigenstrain: 
\begin{equation}\label{eq::eigen}
    \varepsilon_{ij}^*(\varphi) = \eta \, \varphi \, \delta_{ij} \, ,
\end{equation}
where $\eta$ is the lattice mismatch between the two phases and a simple linear approximation for the dependence of $\varepsilon_{ij}^*$ from $\varphi$ has been used. Notice, also, that this choice explicitly disregards the possible presence of other elastic contributions, such as dislocations and grain boundaries, which are therefore not considered in the present study. Since the time-scale of strain relaxation is much faster than the diffusive one of the phase separation, it is customary \cite{khachaturyan2008theory} to assume that mechanical equilibrium holds at any time so that the strain field $\boldsymbol{\varepsilon}$ can be directly computed by solving the (static) equilibrium condition. For a periodic and elastically homogeneous material, this can be conveniently achieved in Fourier space. Following Refs.~\cite{mura1987, khachaturyan2008theory,fratzl_modeling_1999}, the Fourier transform of the equilibrium stress $\boldsymbol{\widehat\sigma}$ field can be directly obtained from the Fourier-transformed order parameter $\widehat\varphi$ as:
\begin{align}\label{eq::stress}
    & \widehat{\boldsymbol{\sigma}} = - \eta \boldsymbol{B}(\vec{q}) \widehat \varphi \, ,\\
    & B_{ij}(\vec{q}) = C_{ijkl} (\delta_{kl} - \Omega_{km} C_{mnop} \delta_{op} q_n q_l) \, ,
\end{align}
where $\vec{q}$ is the Fourier component wavevector and $\Omega_{kn}(\vec{q})$ is the inverse of the Green tensor for acoustic displacement, i.e., $\Omega^{-1}_{kn}= C_{kmno} q_m q_o$.

Thanks to this assumption of quasi-equilibrium, the free-energy functional \eqref{eq::FreeEnergy} just depends parametrically on strain, so that the dynamics of spinodal decomposition can be fully described by the well-known Cahn-Hilliard equation \cite{cahn_free_1958, cahn_phase_1965,garcke2006} as 
\begin{equation}\label{eq::pf_cahnhilliard}
    \frac{\partial \varphi}{\partial t} = M\nabla^2 \mu= M \nabla^2 \left( - \epsilon \nabla^2 \varphi + w'(\varphi)+\rho'(\varphi,\boldsymbol{\varepsilon}) \right) \, ,
\end{equation}
where $\mu=\delta F/\delta \varphi$ is the local chemical potential, $M$ is a mobility constant and the $'$ symbol indicates derivation with respect to the $\varphi$ variable. From eqs.~\eqref{eq::elastic} and \eqref{eq::eigen}, for elastically homogeneous materials, it can be found that $\rho'(\varphi,\boldsymbol{\varepsilon})=-\eta \text{tr}(\boldsymbol{\sigma})$, being $\text{tr}(\boldsymbol{\sigma})$ the trace of the stress tensor set from Eq.~\eqref{eq::stress}. Numerically, the time-integration of \eqref{eq::pf_cahnhilliard} is here conveniently achieved by using a semi-implicit spectral scheme \cite{wang_kinetics_1993} as
\begin{equation}\label{eq::spectral}
    \widehat{\varphi}_{t+\delta t} = \frac{\widehat{\varphi}_t - \delta t M \,q^2 \widehat{w'(\varphi_t)}} {1 + \delta t M \left(\epsilon q^4 + \eta^2 \text{tr}(  \boldsymbol{B}) q^2\right)} \, ,
\end{equation}
where $\widehat{w'}$ indicates the Fourier transform of the first derivative of the double-well potential and $\delta t$ is the integration time-step. Spatial discretization is instead set on a uniform two-dimensional square grid.

From a linear-stability analysis, it can be demonstrated \cite{fratzl_modeling_1999} that the mixed phase is unstable and undergoes spontaneous phase separation when the following criterion is met:
\begin{equation}\label{eq::pf_stability}
    w''(\varphi) + \eta^2 \min_{\vec{q}}\left(\text{tr}(\boldsymbol{B}(\vec{q}))\right)< 0 
\end{equation}
In the strain-free case, the stability of the homogeneous mixture is determined by the curvature of $w(\varphi)$, which indeed defines the spinodal lines in the $(\varphi,T)$-phase diagram. Elastic strain, on the other hand, always penalizes spinodal decomposition, thus stabilizing the homogeneous phase and extending the metastability within a part of the spinodal region. Moreover, by considering anisotropic materials, $\boldsymbol{B}(\vec{q})$ changes according to the crystallographic orientation, thus making phase separation direction-dependent. In particular, this can lead to decomposition suppression along hard crystallographic directions and enhancement along soft ones, thus resulting in anisotropic phase patterns.

In this work, we focus on the case of cubic symmetry and consider an arbitrary material of high Zener anisotropy ratio 
$Z=2C_{44}/(C_{11}-C_{12})=4$ and a wide misfit range $\eta\in[0, 1.2\%]$. Elastic constants are fixed as $C_{11}=C_{44}=2C_{12}=3\times 10^4$ in the arbitrary units of the double well potential and interface term. All values have been chosen to maximize the variability in the phase-separation patterns. By this choice, softer directions are aligned with the $\left\langle10\right\rangle$ and $\left\langle01\right\rangle$ axes so that, for high strains, striped domains are expected to form along them\cite{cahn_cubicACTAMETAL1962,nishimoriPRB1990}.

Fig.~\ref{fig::data}(a) provides a representative example of a PF time-evolution for a moderate misfit value of $\eta \approx 0.482\%$, serving to visually illustrate the type of microstructural transformation considered in this work. As discussed, domains aligned with the $\left\langle 10\right\rangle$ elastically soft directions rapidly emerge from the initial random configuration, although the elastic contributions are not strong enough to prevent the formation of more rounded inclusions. A collection of microstructures, obtained by additional evolutions for varying combinations of $\langle \varphi \rangle$ and $\eta$, is shown in panel (b). Consistent with theory, configurations having $(\langle \varphi \rangle, \eta)$ values outside of the spinodal region (orange region) are stable against small fluctuations around the mean composition, thus remaining in the uniform, mixed state.

\subsection{Dataset generation} \label{sec::methods::data}
A dataset of $2000$ time-sequences of spinodal decomposition has been constructed by the numerical integration of equation \eqref{eq::spectral} for different initial random configurations, average composition $\left\langle \varphi\right\rangle$, and misfit strain $\eta$. All simulations are performed on a $128 \times 128$ uniform square grid of collocation points with periodic boundary conditions naturally enforced by the spectral description. To ensure a smooth resolution of the PF interface, we set $\epsilon=5$, using grid point distance as the unit for length. Since the mobility $M$ just acts as a scaling factor for the time scale (see Eq.~\eqref{eq::pf_cahnhilliard}), we take it unitary and use a time-step $\delta t=0.01$ to achieve a stable numerical solution.

Each sequence in the dataset comprises 100 snapshots, taken at fixed time intervals $\tau$, corresponding to 100 integration time-steps $\delta t$, in such a way that consecutive frames are sufficiently different to provide effective information for the NN training. In the following, we use $\tau$ as the unit of time.

The initial configuration for each simulation is designed to represent a homogeneous mixed phase with local composition fluctuations, eventually seeding the phase separation process and resulting in the typical microstructural evolution patterns reported in experimental works such as \cite{vos_relationship_1966, ardell_modulated_1966}. This is obtained by initializing the composition field in the form of Perlin noise \cite{perlin_image_1985}, a gradient noise algorithm that generates structured correlated random profiles. Unlike white noise, Perlin noise consists of smoother patterns with features that the NN can properly distinguish from actual numerical noise.

In order to achieve robust generalization during model training, a sufficiently diverse set of initial configurations is needed. To this goal, a randomized shifting and rescaling of the Perlin noise profiles is implemented so to vary the average composition $\left\langle\varphi\right\rangle$ in the range $[0.2,0.8]$ (with the present parameters, $\varphi \approx 0.21$ and $\varphi \approx 0.79$ are the limiting compositions for spinodal decomposition at zero-misfit, as shown in the phase diagram of Fig.~\ref{fig::data}(b) and control the extent of the fluctuations. To this latter goal, the maximum deviation of the Perlin noise fluctuations around the average value is randomly sampled from a normal distribution with a standard deviation equal to $0.1$. 

\subsection{CRNN architecture} \label{sec::methods::nn_archi}
In this work, we develop two CRNN architectures similar to the ones proposed in Refs.~\cite{lanzoni_morphological_2022, lanzoni_extreme_2024}, but specialized for the two distinct tasks: (1) NN$_\text{Par}$ for the $\eta$ parameter evaluation, (2) NN$_\text{Evo}$ for the prediction of the time evolution sequence.

\begin{figure}[t!]
 \centering
   \includegraphics[width=0.9\linewidth]{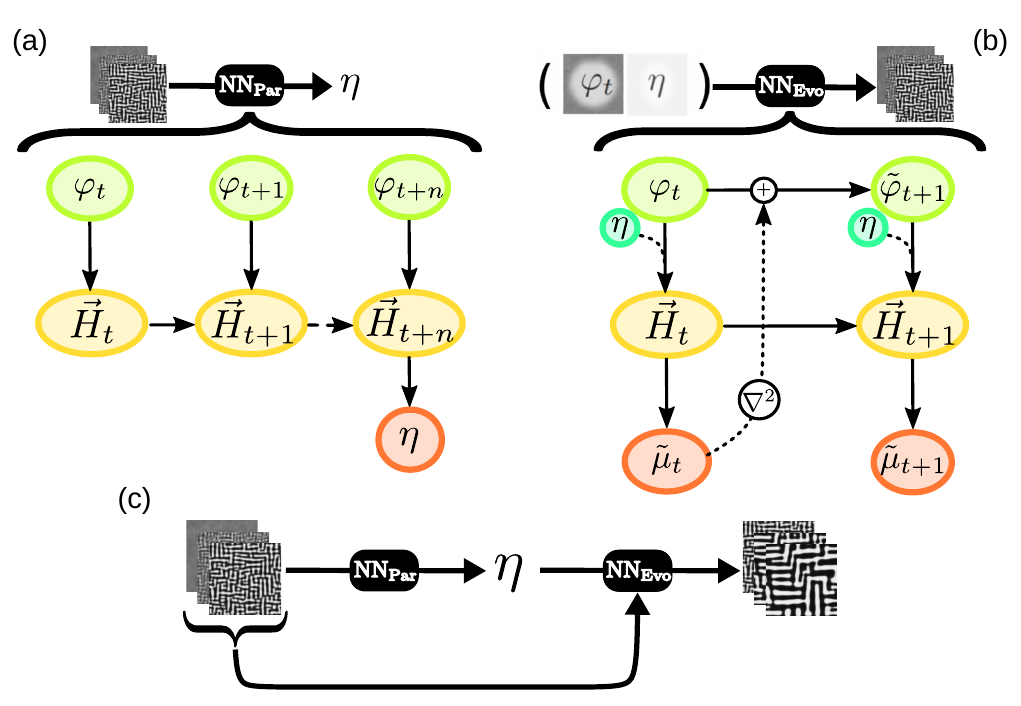}
    \caption{(a) Simplified structure of the NN for parameter extraction (NN$_{\text{Par}}$), showing how the model processes an entire sequence \( \{\varphi_t, \,..., \, \varphi_{t+n}\}\) to infer the lattice mismatch parameter \(\eta\) from the final hidden state. (b) Simplified architecture of the NN for evolution prediction (NN$_{\text{Evo}}$). At each time step, the model predicts the time-evolved state \(\tilde\varphi_{t+\tau}\) from the previous state \(\varphi_t\), a hidden state \(\vec H_t\), and the parameter \(\eta\) provided as a constant field. Rather than predicting \(\tilde\varphi_{t+\tau}\) directly, the network outputs an auxiliary scalar field \( \widetilde{\mu}_t \), from which the updated state is computed. (c) NN$_{\text{Par}}$ and NN$_{\text{Evo}}$ are combined into a cascaded neural network, where the output of one network serves as the input for the next, enabling the possibility of predicting future evolution from a given sequence.} \label{fig::nn_archi}
\end{figure}

In both cases, the CRNN core structure takes as input the phase field at each time step $\varphi_t$, in the form of a sequence of $128 \times 128$ pixel images, and processes it through two stacked convolutional gated recurrent unit (convGRU) blocks~\cite{chung_empirical_2014}. Each convGRU block employs 24 channels and $5 \times 5$ convolutional kernels with circular padding \cite{circular_padding} to enforce periodic boundary conditions by construction, consistently with the dataset examples. These hyperparameters were selected based on insights gained from preliminary testing to ensure optimal performance and stability of the model. Depending on the task considered, the output of each NN is calculated with a specialized subnetwork, as discussed in the following.

While both NN$_\text{Par}$ and NN$_\text{Evo}$ share this overall common architecture, their internal parameters are independently optimized through separate training procedures tailored to their respective learning objectives.

\subsubsection{NN structure for parameter extraction}
The NN for parameter extraction (NN$_\text{Par}$) is derived from the general CRNN architecture discussed before, with the scope of analysing an evolution sequence of spinodal decomposition and predicting the value of the misfit $\eta$ generating it. A simplified sketch of its structure is shown in Fig.~\ref{fig::nn_archi}(a). In particular, the NN$_\text{Par}$ takes a sequence of consecutive time frames as input, processes it through the convGRU layers, and then, through the final layer, transforms the last hidden state into the predicted scalar parameter $\eta$, using strided convolutions as a pooling strategy. Since NN$_\text{Par}$ only uses convolutions and pointwise functions, it can in principle be applied to inputs of larger size. Notice, however, that the prediction may no longer be a single scalar. For example, the application to a $256 \times 256$ computational domain would yield a $2 \times 2$ output, which may be heuristically interpreted as the predictions for the domain quadrants. A single value for $\eta$ may then be obtained by taking the average of these values. Similar considerations hold for larger computational domains.

In NN$_{\text{Par}}$'s training procedure, the model is provided with 100 snapshots from a sequence and is required to provide the misfit parameter only after the last one. The standard Mean Squared Error (MSE) loss is used:
\begin{equation} \label{eq::loss-par}
    L (\theta_\text{Par}) = \frac{1}{N_{\text{ts}}} \sum_{i=1}^{N_{\text{ts}}} (\eta_i - \widetilde{\eta}_i)^2
\end{equation}
where $\widetilde{\eta}_i$ is the predicted misfit for each case $i$ of the $N_\text{ts}$ cases in the training set and $\theta_\text{Par}$ represents the set of the NN parameters.

Since the inference of the misfit parameter relies on the recognition of the peculiar morphological features observed during phase separation, the training of NN$_\text{Par}$ is performed on the subset of $\approx 1500$ sequences, divided in a 4:1 proportion between training and validation set, obtained from the full dataset after excluding those cases in which the initial small perturbations rapidly decay yielding an homogeneous configuration and, as such, are weakly dependent on $\eta$. Including such sequences in the dataset would therefore provide no meaningful information to the network, resulting in worse predictive performance. This is expected, as it is impossible to uniquely map a uniform, stationary $\varphi$ to the corresponding $\eta$.
Moreover, data augmentation techniques, including reflections, 90$^\circ$ rotations, and the $\mathbb{Z}_2$ symmetry transformation $\varphi \rightarrow (1-\varphi)$, are employed to further improve the robustness and diversity of the training dataset. 

Training of the NN parameters is performed using the standard implementation of the Adam optimizer \cite{kingma_adam_2014}. Once trained, the model can accept input sequences of any length, which would be beneficial especially in the perspective analysis of, generally scarce and costly, experimental data.

\subsubsection{NN structure for evolution prediction} \label{sec::nn_evo}
The NN$_\text{Evo}$ is developed from the above-discussed CRNN framework to take a configuration $\varphi_0$ (or a short series of frames) and a provided misfit parameter as inputs and then returning the predicted time-evolved sequence as output.

To explicitly handle the parameter $\eta$ into the NN inputs, its value is simply concatenated to $\varphi_t$, i.e., the couple $(\varphi_t, \eta)$ is internally represented as a single $128 \times 128$ image with two channels. Despite its simplicity, this approach is particularly convenient for its ease of implementation. Moreover, it does not break the fully-convolutional character of NN$_\text{Evo}$, allowing for generalization to arbitrary domain sizes. A simplified sketch of the NN$_{\text{Evo}}$ structure is shown in Fig.~\ref{fig::nn_archi}b.

As already discussed in Ref.~\cite{lanzoni_extreme_2024}, the basic CRNN architecture lacks any constraint on the conservation of $\varphi$. A significant improvement in the consistency of the NN predictions with the conservative flow dynamics prescribed by Eq.~\eqref{eq::pf_cahnhilliard}, especially for time-extrapolation, is achieved by introducing an additional physics-inspired output layer, reflecting the formulation of the CH equation. Rather than predicting the next state of the field $\widetilde{\varphi}_{t+\tau}$ directly (tilde distinguishing again predicted quantities from dataset ones), the network is designed to output the scalar field $\widetilde{\mu}_t$, taking the role of the local chemical potential (multiplied by $M$) in Eq.~\eqref{eq::pf_cahnhilliard} so that the new frame can be generated as:
\begin{equation}
    \widetilde{\varphi}_{t+\tau} = \widetilde{\varphi}_t + \nabla^2 \widetilde{\mu}_t
\end{equation}
This is an alternative version of the approach implemented in Ref.~\cite{lanzoni_extreme_2024}, where the flow current $\vec{J}=-M\nabla\mu$ was evaluated to enforce exact conservation of $\varphi$ by construction and is here found to provide superior results in the case at hand.

In order to train the NN$_\text{Evo}$ model, we define the loss function as the spatial and temporal average of the weighted sum of three components: the mean squared error (MSE) between the ground truth and predicted sequences ($L_{\phi}$), the MSE between the squared gradients of these sequences ($L_{\nabla}$), and the MSE between the ground truth double well energy and the one computed from the predicted sequences ($L_{W}$):
\begin{equation} \label{eq::loss-evo}
    L (\theta_\text{Evo}) = \frac{1}{N_{\text{ts}}T} \sum_{i=1}^{N_{\text{ts}}} \sum_{t=1}^{T} \left ( \langle 
    L_{\phi}(\theta_\text{Evo}) + 
    \lambda_{\nabla} L_{\nabla}(\theta_\text{Evo}) \rangle  + 
    \lambda_{W} L_{W}(\theta_\text{Evo}) \right )
\end{equation} 
with
\begin{equation}
    \begin{aligned}
    L_{\phi}(\theta_\text{Evo}) & = (\varphi_i(t) - \widetilde{\varphi}_i(t|\theta_\text{Evo}) )^2 \\ 
    L_{\nabla}(\theta_\text{Evo}) & = (|\nabla \varphi_i(t)|^2 - |\nabla \widetilde{\varphi}_i(t|\theta_\text{Evo})|^2)^2 \\
    L_{W}(\theta_\text{Evo}) & = (\langle w(\varphi(t))\rangle- \langle w(\widetilde{\varphi}(t|\theta_\text{Evo})) \rangle)^2
    \end{aligned}
\end{equation}
Here, $\theta_\text{Evo}$ represents the set of NN parameters, $i$ indexes the $N_{\text{ts}}$ elements of the training set, $t$ represents the time step, ranging from 1 to the total sequence length $T$, and $\langle \, . \, \rangle$ indicates the spatial average. The parameters $\lambda_{\nabla}$ and $\lambda_{W}$ are the two weights controlling the relative importance of the gradient and double well energy terms, respectively. Based on preliminary analysis, which evaluated the impact of these terms in the loss function on the temporally evolved profiles predicted by the model, they were set to 60 and 150, respectively.

While the $L_{\phi}(\theta_\text{Evo})$ term evaluates the pixel-by-pixel correspondence between the predicted and true fields, $L_{\nabla}(\theta_\text{Evo})$ is included as a local regularization term on the gradient of the predicted field, penalizing oscillations in internal areas of the composition domains. $L_{W}(\theta_\text{Evo})$, on the other hand, is justified by the fact that the spatial integral of the double well term is related to the total interface length between the two phases~\cite{salvalaglioCGD2015}, hence it penalizes on a global level predictions involving broader or smaller domain boundaries. In our tests, these terms indeed help to eliminate small artifacts that can emerge during the evolution prediction.

The full dataset of $2000$ simulations is exploited for the training of NN$_\text{Evo}$, with a 4:1 random partitioning between training and validation. The same data augmentation techniques used for NN$_\text{Par}$ are exploited for the training of this model as well. Additionally, to enhance the model's generalization capabilities, noise injection regularization is applied during training by introducing small random perturbations such that $\varphi \rightarrow \varphi + \delta$, where $\delta$ is sampled from a Gaussian white noise, as in our previous work in Ref.~\cite{lanzoni_extreme_2024}. Also in this case, the standard implementation of the Adam optimizer \cite{kingma_adam_2014} is used to perform the training of the NN parameters. 

In NN$_{\text{Evo}}$'s training procedure, the so-called curriculum learning technique \cite{bengio_curriculum_2009} is used. In the initial epoch, the model is provided with 99 snapshots from a sequence and tasked with predicting only the final step, with the loss evaluated solely on that last prediction. As training progresses, the number of input snapshots is gradually reduced, which requires the model to predict increasingly longer portions of the sequence. By the end of the ramp, the model has only the initial snapshot and must predict the remaining sequence, with the loss evaluated on all of its predictions. Although this procedure may lead to an apparent initial increase in loss, it facilitates more stable and efficient learning by progressively introducing the complexity of the task.

Once trained, the model offers flexibility in both spatial and temporal dimensions: its fully convolutional architecture enables it to process input images of arbitrary size, while the recurrent structure allows for the generation of sequences of any desired length.

As a performance figure of merit, the NN$_\text{Evo}$ model achieves a speed-up of approximately $40 \times$ on a $128 \times 128$ domain, $90 \times$ on a $256 \times 256$ domain, and $145 \times$ on a $512 \times 512$ domain compared to the semi-implicit numerical scheme used to generate the dataset. Notably, the machine learning approach scales linearly with the number of collocation points~\cite{lanzoni_extreme_2024}, making it especially well-suited for simulations on large domains.

\subsubsection{Combining the models in the cascade NN approach}
By combining the two trained models into a cascaded neural network, where the output of NN$_\text{Par}$ is fed as input to NN$_\text{Evo}$, it becomes possible to predict the future evolution of a system directly from an observed sequence. A sketch of this workflow is reported in Fig.~\ref{fig::nn_archi}(c). At variance with the stand-alone usage of NN$_\text{Evo}$, in this modality it is possible to input the entire evolution sequence already used in NN$_\text{Par}$ instead of the single last frame, thus operating NN$_\text{Evo}$ in sequence-to-sequence mode, possibly yielding better conditioning.

Aside from the already discussed higher degree of transparency, this cascaded approach also preserves the possibility of partly re-using architectures should more effective or efficient modules become available: since NN$_\text{Par}$ and NN$_\text{Evo}$ are separated, they can be more easily and independently replaced with respect to ``monolithic'' approaches.

\section{Results} \label{sec::results}
\begin{figure}[t!]
 \centering
   \includegraphics[width=1\linewidth]{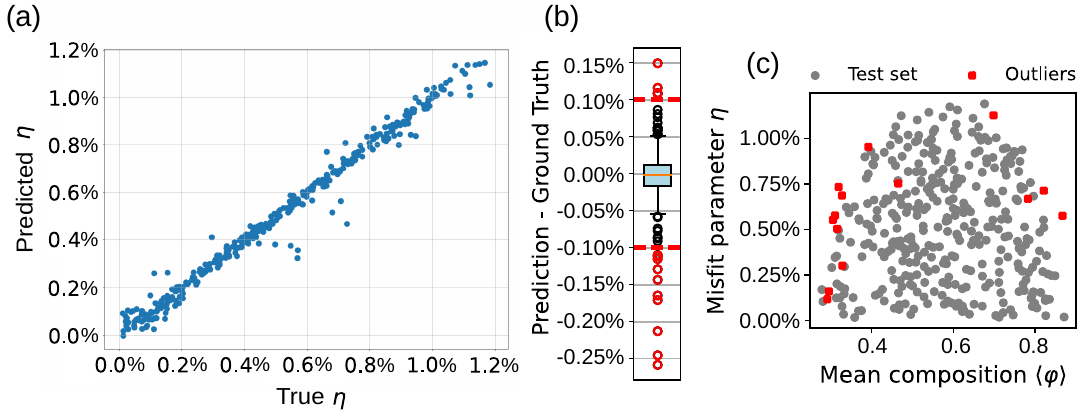}
    \caption{(a) Misfit parameter $\eta$ regression plot for 359 test sequences exhibiting phase separation, showing strong correlation and high predictive accuracy of the NN$_{\text{Par}}$ model.
    (b) Boxplot of prediction errors, with most $\eta$ values predicted within a strain error of $\approx 0.013\%$. The red dashed lines are placed to highlight the few outliers exceeding a $0.1\%$ absolute misfit error. We recall that the light blue box delimits the first and third quartiles, while the whiskers are here $1.5\times$ the interquartile range.
    (c) Distribution of true $\eta$ and mean composition $\langle \varphi \rangle$ for the test set. Outliers (in red) are scattered along the spinodal boundary, where phase separation is slower and less distinct.} \label{fig::fig-par1}
\end{figure}

We here discuss the CRNN performances following the logic flow of its final usage, i.e. first extracting from existing time-evolutions (from PF simulations) the misfit parameter (Sect.~\ref{sec::results::para}), then using explicitly the known parameter to predict a desired evolution sequence (Sect.~\ref{sec::results::evo}). Once the individual models have been validated, we merge the two steps into the cascade approach (Sect.~\ref{sec::results::cascade}).

\subsection{Parameter extraction} \label{sec::results::para}

To evaluate the predictive performance of the NN$_{\text{Par}}$ model, we constructed a test set consisting of 500 sequences generated under the same conditions as the original dataset discussed in \ref{sec::methods::data}. This set was subsequently filtered to retain only the 359 sequences that exhibited phase separation, excluding those where no separation occurred.

Fig.~\ref{fig::fig-par1}(a) presents the regression plot of predicted versus true $\eta$ values, obtained using the best-performing model on the validation set. For each evaluation, the NN is fed sequences composed of 100 snapshots. The strong correlation, with an $R^2$ value of 0.983, highlights the model’s ability to accurately infer the parameter governing phase evolution. Most $\eta$ values are predicted within an absolute error of approximately $0.013\%$, as indicated by the boxplot in Fig.~\ref{fig::fig-par1}(b), the light blue box indicating the boundaries of the first and third quartile, i.e. enclosing $50\%$ of data. Some noticeable outliers, with an absolute error exceeding $0.1\%$,  are nonetheless present, suggesting that certain conditions pose challenges for the model's predictions. Their nature can be better understood by examining the corresponding true $\eta$ values and mean compositions $\langle \varphi \rangle$. By highlighting in red the most significant ones in a misfit vs average $\varphi$ plot, it becomes apparent that the outliers are located at the outskirts of the spinodal region, as illustrated in Fig.~\ref{fig::fig-par1}(c), i.e., where the competition between phase separation and the homogeneous mixing is more critical. Moreover, in these parameter regions, spinodal decomposition occurs slowly even when it happens. During the 100 snapshots of the input sequence, the process remains in its early stages, showing only minimal differences in the $\eta$-dependent morphologies. This lack of features poses a significant challenge for the model in accurately predicting $\eta$ for such critical cases.

\begin{figure}[t!]
 \centering
   \includegraphics[width=0.9\linewidth]{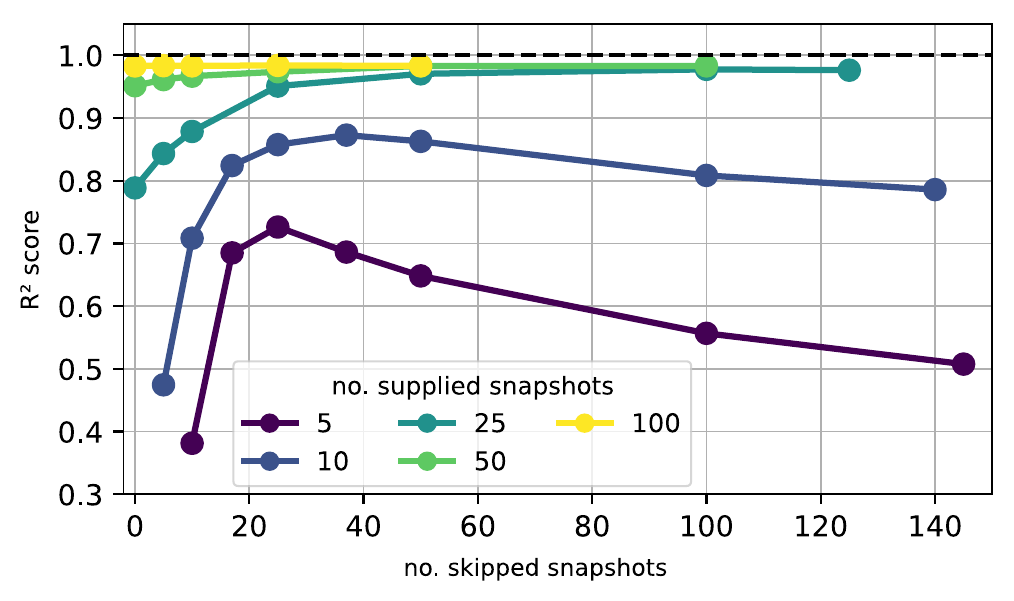}
    \caption{Prediction accuracy (measured via $R^2$) as a function of the number of supplied snapshots and the number of skipped frames from the initial Perlin noise profiles. Using 100 snapshots consistently yields high accuracy, regardless of the starting point of the supplied snapshots. Using just 25 snapshots, a similar high accuracy can be reached. For shorter sequences, predictive performance improves significantly when early stages of phase separation are skipped.} \label{fig::fig-par2}
\end{figure}

An important figure of merit for the NN predictive capability is the minimum length of the time sequence required to obtain reliable estimations of the misfit. While NN$_{\text{Par}}$ was trained to predict $\eta$ after processing sequences of 100 snapshots, the model can be identically applied to sequences of any length. We therefore investigate the relationship between the number of input snapshots and the model's predictive performance on the test set by truncating the sequences at different times. This has been done either by dropping the initial part of the dynamics, by removing the final configurations, or by extracting an internal subsequence, so to assess the impact of the different evolution stages on the NN performances. In particular, since $\eta$-dependent morphological differences are more subtle during the early stages of phase separation, we expect higher accuracy for input sequences starting from already partially separated configurations after a few $\tau$. Moreover, differences should be amplified when shorter dynamics are used. Due to the non-linear nature of NN, however, a thorough evaluation of all these aspects is necessary. In Fig.~\ref{fig::fig-par2}, we analyse the $R^2$ coefficient as a function of both the number of snapshots separating the initial one provided from Perlin noise and the total sequence length. As expected, the highest accuracy is achieved when providing sequences of 100 snapshots, with no substantial difference observed whether the sequences begin from the initial Perlin noise or from later, well-separated stages, such as after $50\,\tau$. When the number of snapshots is reduced to 50, predictions remain robust: supplying dynamics evolving from later stages, thereby skipping the initial transient of phase separation, yields accuracy comparable to that of the full-length input. With only 25 snapshots, however, the model shows a drop in performance when sequences start from the mixed state. Notably, predictions reach the same level of accuracy given by supplying 100 snapshots, if these 25 frames are taken from later stages (at least from $50\,\tau$), suggesting that the model benefits from input sequences reflecting clearer morphological differentiation. For shorter inputs of 10 or 5 snapshots, predictive accuracy deteriorates significantly, even when initial transient dynamics are skipped. Interestingly, for these extremely short inputs, a narrow optimal window exists: prediction improves slightly when skipping 25–50 frames for the 10-snapshot case and 20–30 frames for the 5-snapshot case. Beyond these points, accuracy declines again, likely due to the decreasing differences between subsequent frames in more evolved phase-separated morphologies, resulting in less information for the NN.

\subsection{Evolution prediction} \label{sec::results::evo}

\begin{figure}[t!]
 \centering
   \includegraphics[width=0.9\linewidth]{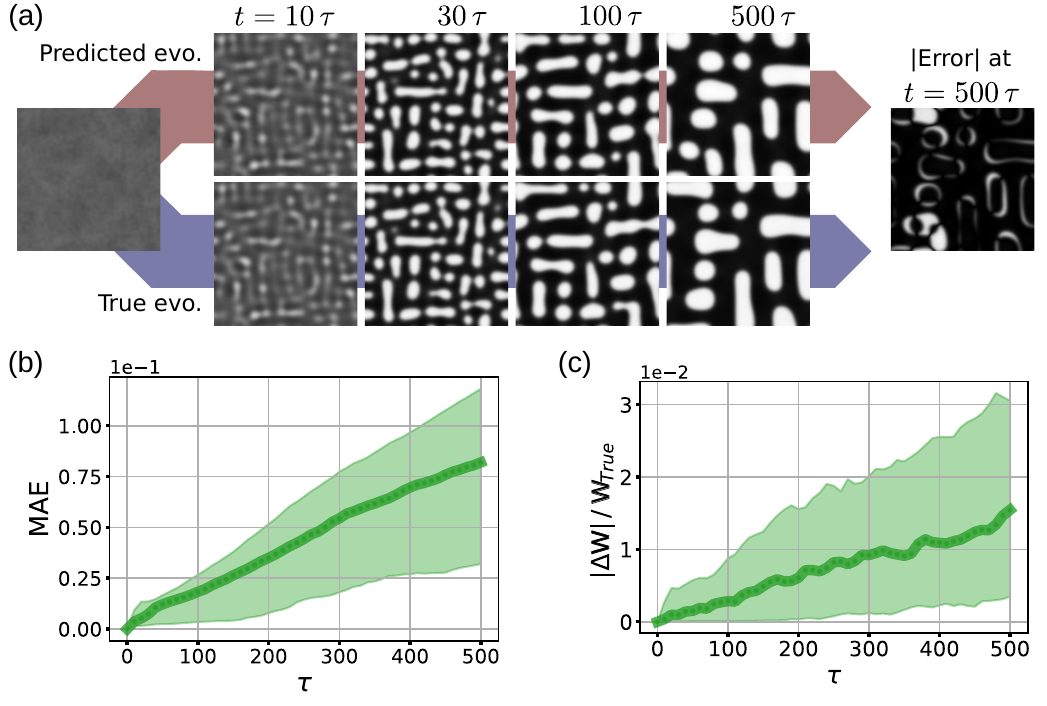}
    \caption{(a) Test sequence of spinodal decomposition with $\eta\approx0.45\%$, comparing the NN$_\text{Evo}$ prediction with the ground-truth PF evolution. The absolute error across the domain, evaluated as the pointwise difference between the two profiles, is also shown for the last frame. (b,c) Time evolution of (b) MAE and (c) relative interface length difference $|\Delta W|/W$ between NN and true models obtained on a test set of 500 sequences, extended to $500\,\tau$. The median value (solid line) and interquartile range (shaded region) are shown.} \label{fig::fig-evonew}
\end{figure}

Following the training and validation process, we evaluated the NN$_{\text{Evo}}$ model's predictive capability on an independent test set to ensure robustness and generalizability. For this purpose, we selected the model that achieved the best performance on the validation dataset. The test set comprises 500 sequences generated with the same characteristics as the training dataset but extended to $500\,\tau$ in length, i.e. 5 times longer than the sequences used during training, so as to also evaluate the model's behaviour in time extrapolation. As an illustrative example, in  Fig.~\ref{fig::fig-evonew}(a) we report a comparison between the NN predicted and ground truth dynamics for a sequence with $\eta \approx 0.45\%$ taken at random from the test set. In the initial stages, almost a one-to-one match is achieved. However, due to the accumulation of small prediction errors across the time steps and to critical bifurcation events, such as domain splitting and coalescence, the pixel-level match progressively degrades, and local discrepancies in the morphologies are observed. At time $t=500\,\tau$ (see Fig.~\ref{fig::fig-evonew}(a)), it may be noted that the predicted states are nonetheless still in qualitative accordance with the ground truth.

To provide a quantitative evaluation of the accuracy of the model’s predictions over time, we compute the Mean Absolute Error (MAE) at regular intervals during the evolution. Although the model was trained by minimizing the mean squared error (MSE), the MAE provides a more interpretable metric in this context, as it directly quantifies the average number of incorrectly predicted pixels. In Fig.~\ref{fig::fig-evonew}(b), we present a statistical analysis of the time evolution of MAE evaluated on the whole test set sequences. The median MAE (solid line) and the interquartile range (shaded region), calculated across the whole test set, reveal that the relative error remains generally below $2.5\%$ at $100\,\tau$ and increases to $\approx 8\%$ after $500\,\tau$, similarly to other works in the literature~\cite{alhadaNPJCM2024}. 
It should be noted that the amplification of initial small deviations during evolution is expected, as even minor differences can lead to diverging morphological outcomes. This is not an exclusive issue of the NN model, but can already be observed in PF simulations with varying spatial and temporal discretizations. Then, while MAE is a good metric for the degree of overlap between predicted and true frames, it may be too strict for evaluating how the overall model prediction complies with the underlying physical process, as in such a case, only the average microstructural properties should be regarded.

As an alternative to MAE, a key global variable to consider the model behaviour is the total interface length, which is here conveniently estimated as the integral of the double well $W=\langle w(\varphi)\rangle$~\cite{salvalaglioCGD2015}, following the same approach also used for the regularization term $L_W$ (see Sect.~\ref{sec::nn_evo}). In Fig.~\ref{fig::fig-evonew}(c), we show the trend of the relative interface length difference on the same independent test set, highlighting the median and interquartile range. The relative error remains below \(1\%\) for the first $100\,\tau$, and is generally under \(3\%\) even at $500\,\tau$. Such low values confirm that the NN model retains a good level of physical accuracy even beyond the training regime, despite losing the ability to reproduce one-to-one the true field.

\begin{figure}[t!]
 \centering
   \includegraphics[width=1\linewidth]{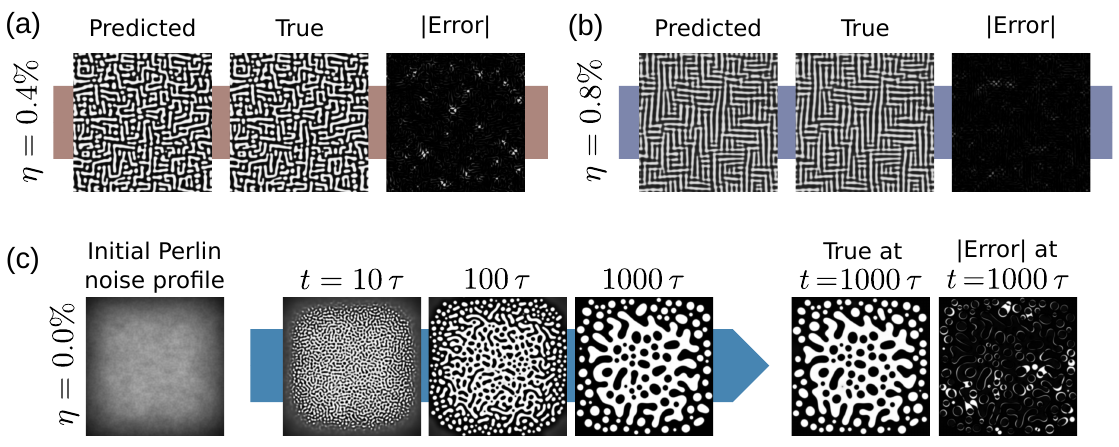}
    \caption{(a,b) Comparison between predicted and true profiles for spinodal decomposition at $t=100\,\tau$, starting from initial Perlin noise profile on large $512 \times 512$ domains, for (a) $\eta = 0.4\%$, and (b) $\eta=0.8\%$. The absolute error map over the domain is also reported. (c) Example of a $1000\,\tau$-long evolution sequence on a $512 \times 512$ domain with $\eta = 0\%$ and starting from a spatially varying initial composition: higher in the center and lower at the edges. Despite local differences, the NN$_\text{Evo}$ prediction is in accordance with the ground truth up to $1000\,\tau$, demonstrating generalization capabilities beyond the training distribution.} \label{fig::fig-evojoint}
\end{figure}

Now that we assessed the predictive capabilities of NN$_\text{Evo}$, we exploit one of the major advantages of using a fully-convolutional architecture and apply it to much larger domains to test its robustness and generalization capabilities beyond the training domain size. Fig.~\ref{fig::fig-evojoint}(a,b) compares NN predictions and PF ground truth morphologies obtained after evolution for $100\,\tau$ on $512 \times 512$ domains (16 times larger than those used during training). Two representative cases with $\langle \varphi \rangle = 0.45$ and $\eta = 0.4\%$ and $0.8\%$ are selected to span the morphological variability. Aside from local discrepancies related to domain pinching, the predicted profiles are in excellent agreement with the true ones, accurately capturing the dominant morphological features and large-scale evolution trends.

To further test the model, we analyse the NN performances in predicting the evolution of peculiar configurations that differ qualitatively from the Perlin noise ones of the training set. As an example, in Fig.~\ref{fig::fig-evojoint}(c), we consider an initial condition featuring a spatially varying average composition, which is higher in the centre and lower toward the edges of the domain. A $512 \times 512$ large domain and a long evolution time of $1000\,\tau$ ($10\times$ the training set) are considered. Despite this deviation from the training distribution, the model accurately captures the overall morphological evolution. This demonstrates strong robustness and generalization capabilities, and the effectiveness of the learning framework in scaling to larger domains.

\begin{figure}[t!]
 \centering
   \includegraphics[width=0.9\linewidth]{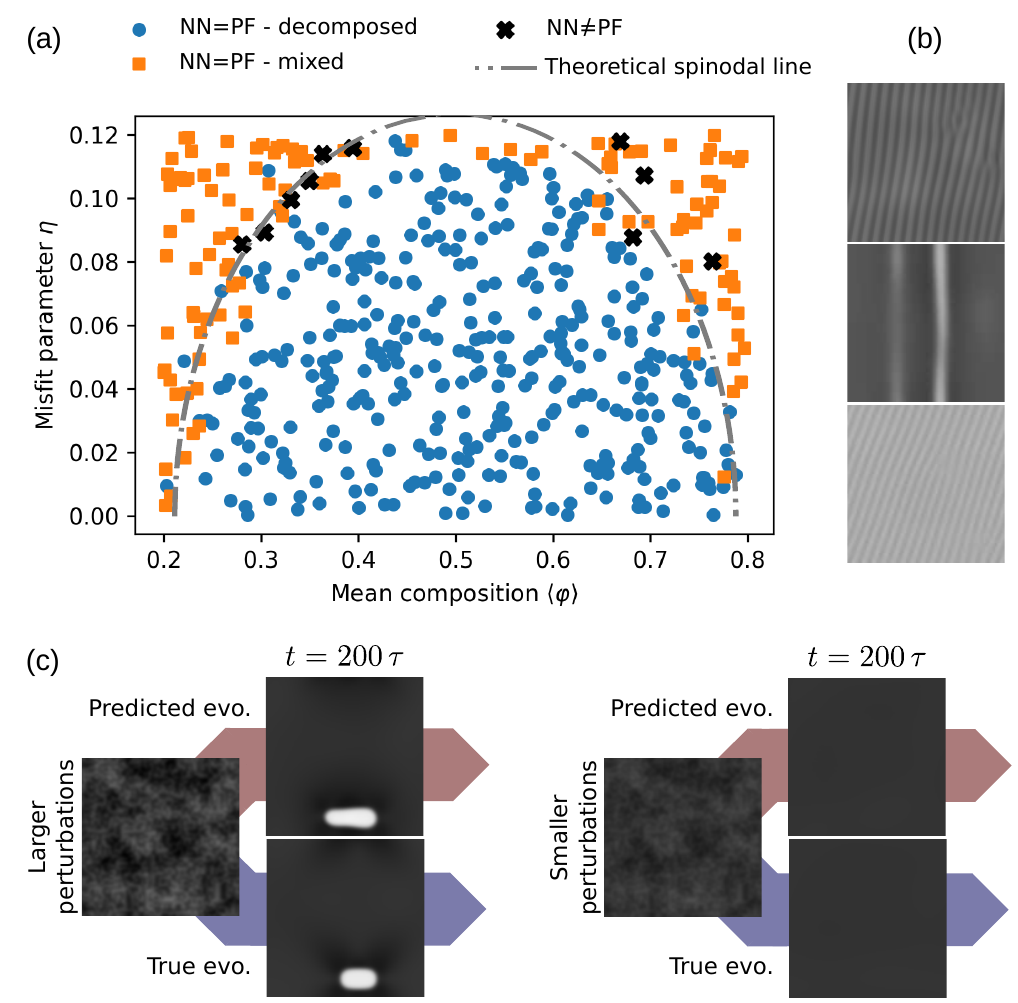}
    \caption{(a) Predicted outcomes across a range of $\langle \varphi \rangle$ and $\eta$ values, illustrating the model’s ability to distinguish between phase separation and uniform mixing. The analytical spinodal line is reported as a dashed gray line for reference. (b) Representative failure cases at $100\,\tau$ where the model incorrectly predicts phase separation or introduces artifacts (black crosses).
    (c) Comparison between predicted and ground truth profiles at $200\,\tau$ for two evolutions starting from the same Perlin noise profile but with different perturbation amplitudes. The one with larger perturbations (left) undergoes phase separation while the other (right) remains mixed. Both cases are captured by the NN model. Simulations are performed on $128 \times 128$ domains.} \label{fig::evo_anal}
\end{figure}

As a final test, we show that the model accuracy is high enough to reconstruct the phase diagram. This is performed by randomly sampling a $(\langle \varphi \rangle, \eta)$ couple and evolving a corresponding Perlin-noise field for 100 steps, after which the amplitude of fluctuations is monitored. If the variation between the highest and the lowest values of the field is lower than 0.05, we label the configuration as mixed; otherwise, we take it as decomposed. The results of this analysis obtained from NN are then compared with the ground truth prediction from PF simulations under the same conditions. A computational domain of $128 \times 128$ has been used in this case. Results are reported in Fig.~\ref{fig::evo_anal}(a). The model predicts the conditions under which phase separation occurs consistently with the ground truth (blue circles). Similarly, it also returns a flat homogeneous mixture when \(\langle \varphi \rangle\) is sufficiently imbalanced to inhibit the spinodal decomposition process (orange squares) in most situations. This demonstrates the ability of the model to implicitly learn the role that elastic strain plays in penalizing spinodal decomposition.

Notice that $\langle \varphi \rangle$ and $\eta$ alone are insufficient to unambiguously determine the onset of phase separation in PF simulations, as the amplitude of the initial perturbations also plays a significant role. Indeed, as shown in Fig.~\ref{fig::evo_anal}(a), an overlap between the spinodal decomposition regime and the region where initial profiles evolve toward uniform profiles can be seen, instead of a sharp boundary that separates the two regimes, which nonetheless is present both for ground truth and NN evolutions.

In a few configurations (10 out of 500), the model was unable to correctly predict the outcome of the dynamics (black crosses). Fig.~\ref{fig::evo_anal}(b) reports three representative examples where NN$_\text{Evo}$ fails to reproduce the correct behavior at time $100\,\tau$. In these cases, the true evolutions remain uniform, whereas the model incorrectly predicts phase separation or introduces spurious artifacts and unphysical features. Such artifacts are not observed in other examples, except in critical conditions where the system lies near the threshold between phase separation and uniform mixing. No cases in which the NN wrongly predicts a final uniform composition state are observed.

To further illustrate the influence of the initial perturbation amplitude on whether the system undergoes phase separation or remains mixed, Fig.~\ref{fig::evo_anal}(c) presents two evolutions starting from the same Perlin noise profile but with different perturbation amplitudes. While the configuration with larger perturbations undergoes spinodal decomposition, the other evolves to a uniform field. The NN model is found to accurately capture both outcomes, highlighting its sensitivity to the initial perturbation even for borderline conditions between phase separation and homogeneous phase.

\begin{figure}[b!]
 \centering
   \includegraphics[width=1\linewidth]{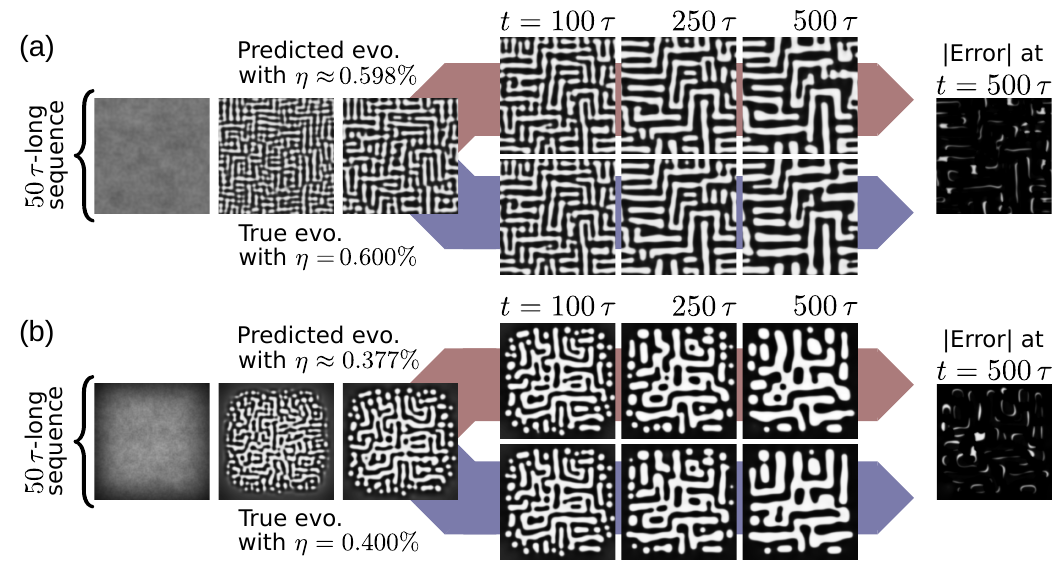}
    \caption{(a) A sequence of 50 snapshots on a $256 \times 256$ domain is processed by NN$_{\text{Par}}$, which estimates \(\eta\) as $0.598\%$, remarkably close to the true value of $0.600\%$. This estimated parameter is then used by NN$_{\text{Evo}}$ to predict the system’s evolution up to $500\,\tau$, showing strong agreement with the true phase-field dynamics. (b) A sequence with spatially varying average composition (higher in the center, lower at the edges) on a $256 \times 256$ domain. NN$_\text{Par}$ predicts $\eta \approx 0.377\%$ (true value: $0.400\%$), and NN$_\text{Evo}$ accurately reproduces the system's evolution up to $500\,\tau$.} \label{fig::chain}
\end{figure}

\subsection{Cascade approach} \label{sec::results::cascade}

Having thoroughly assessed the predictive capabilities of both NN$_\text{Par}$ and NN$_\text{Evo}$ models, we now show how the latter can be used as a downstream module with respect to the former to jointly infer the misfit parameter from a sequence and then, using such explicit knowledge, provide an accelerated simulation of its dynamics. This approach is particularly significant, as it demonstrates the potential for future implementations to analyse experimentally recorded temporal evolutions and predict the system's temporal dynamics without prior knowledge of the underlying global parameter, in this case, $\eta$.

\begin{figure}[b!]
 \centering
   \includegraphics[width=0.9\linewidth]{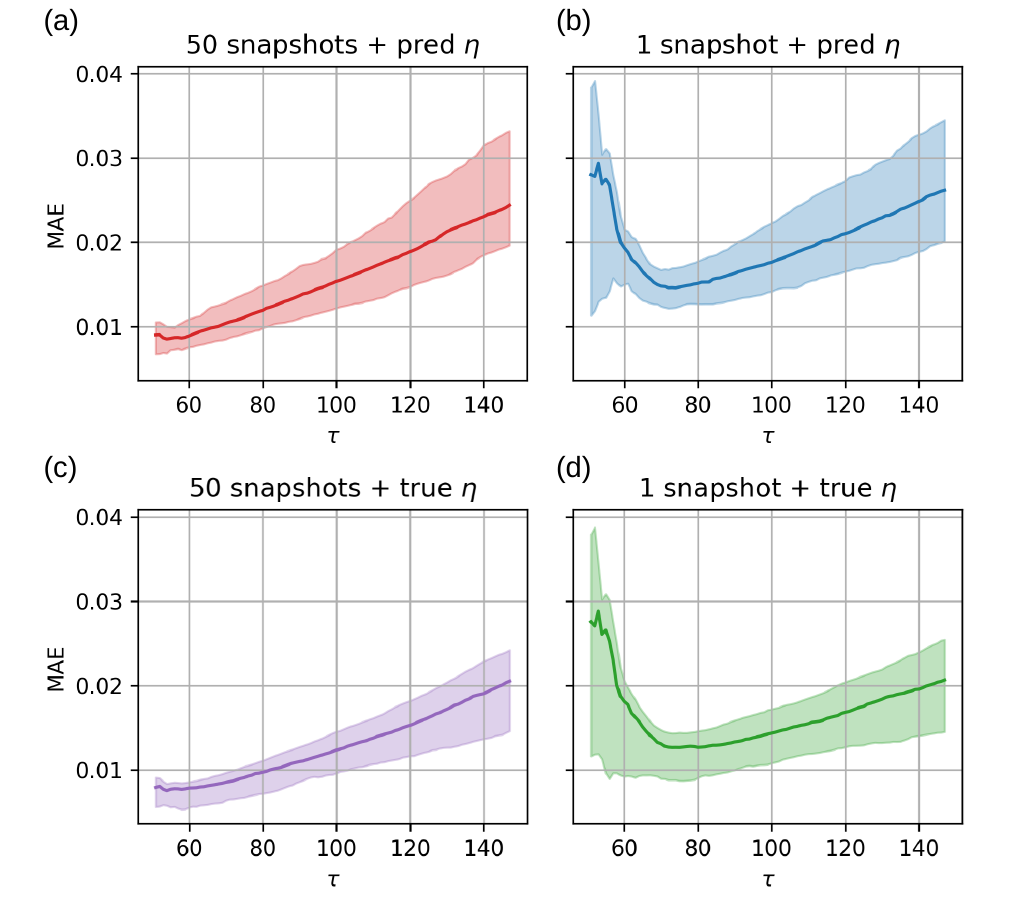}
    \caption{Median and interquartile range of MAE over time for different NN$_{\text{Evo}}$ input configurations: (a)  full 50-snapshot sequence and 3$\eta$ predicted by NN$_{\text{Par}}$; (b) single frame at $50\,\tau$ and predicted $\eta$; (c)  full 50-snapshot sequence and true $\eta$; (d) single frame at $50\,\tau$ and true $\eta$.
    } \label{fig::seq-mae}
\end{figure}

Two examples of this procedure are reported in the Fig.~\ref{fig::chain}. In both cases, a 50-snapshot sequence generated by PF on a $256\times256$ domain is first processed by NN$_{\text{Par}}$ to predict $\eta$ and then fed as input, combined with the predicted $\eta$, to NN$_{\text{Evo}}$, which generates the subsequent evolution up to $500\,\tau$. In the case (a), the sequence is set conformal to the ones of the training set as the composition field is initialized by a uniform Perlin noise around the $\langle \varphi \rangle=0.45$ mean value. The parameter is extracted with high accuracy (absolute error $\approx 0.002\%$ misfit), leading to an accurate reproduction of the subsequent evolution stages. Only local deviations, still resulting from critical events of coalescence and splitting, are observed even in the late stage, as made evident by the error map. In the second case of Fig.~\ref{fig::chain} (b), instead, we consider an initial configuration such that Perlin noise fluctuations are stronger in the center of the computational domain and taper off in intensity towards the edges. In this example, the predicted parameter is $\eta \approx 0.377\%$, compared to the ground truth value of $\eta = 0.400\%$. The error of $\approx 0.013\%$ is fully consistent with the previous findings in the analysis of NN$_{\text{Par}}$~\ref{sec::results::para}. Despite such a discrepancy, the predicted evolution still closely resembles, even after $450\,\tau$, the true one, thus giving a measure of fault tolerance.

In the cascade approach, it is also possible to take advantage of the availability of the initial sequence used to predict the parameter to feed a more informative input to NN$_\text{Evo}$. To assess the expected improvement of considering this sequence-to-sequence approach instead of the single-to-sequence one analysed in Sect.~\ref{sec::results::evo}, we compare performances on the 359 sequences dataset also used in Sect.~\ref{sec::results::para}. To this goal, we first supply to NN$_{\text{Par}}$ the initial 50-frames-long sequences, obtaining an estimation for the respective $\eta$ values. Then, we proceed with predicting the subsequent time evolution sequences using NN$_\text{Evo}$ using either the full 50-snapshot sequence or its last frame only. In both cases, we then statistically analyse the MAEs of the prediction by computing their median and interquartile range over time, resulting in the curves of Figs.~\ref{fig::seq-mae}(a) and (b) respectively. Using the 50-frame sequence as input noticeably regularizes the early stages of prediction, leading to improved accuracy. This benefit is, however, limited to the initial stages as the MAE curves converge to the same long-term behaviour, due to small error accumulation and the underlying bifurcating dynamics.

Finally, it remains to disentangle the error due to the evaluation of $\eta$ by NN$_{\text{Par}}$ from the one of predicting the time-sequence by NN$_{\text{Evo}}$. To this goal, we repeat the same analysis of MAEs as in Figs.~\ref{fig::seq-mae}(a) and (b) but supplying the true $\eta$ values into NN$_{\text{Evo}}$. The results are shown in Figs.~\ref{fig::seq-mae}(c) and (d). As expected, the use of the true $\eta$ improves the quality of the predicted evolution. For all cases of Fig.~\ref{fig::seq-mae}, it is, however, clear that neither model can be considered the sole accuracy bottleneck responsible for the observed error accumulation.

\section{Conclusions} \label{sec::conc}
Based on suitable convolutional recurrent NN architectures, we proposed a cascade model able to both learn the strain condition and predict the temporal evolution of alloy microstructures during coherent spinodal decomposition. The model enables an accurate extraction of key physical parameters. In particular, NN$_\text{Par}$ predicts the misfit strain $\eta$ with a mean absolute error of only $0.013\%$. Notably, such small differences lead to almost visually imperceptible effects in the resulting morphologies. This highlights the potential of the approach as a robust inverse modeling tool.

At the same time, the prediction module, NN$_\text{Evo}$, delivers substantial speed-ups if compared to the numerical method used to generate the dataset. Importantly, it has a linear scaling with the number of collocation points, making it particularly well-suited for large-scale simulations. Testing demonstrated high accuracy in sequence prediction despite considering a very wide range of lattice misfit and average compositions. Also, extrapolation to long evolutions (up to 5 times the sampled one in training) was shown to yield consistent results, despite some observed deviations beyond the training regime. Moreover, the fully-convolutional and recurrent architecture of the models was shown to enable application to arbitrary domain sizes and sequence lengths.

The proposed framework is versatile and applicable to the broader class of continuum models beyond the specific Cahn-Hilliard equation considered here, which was chosen as particularly well suited to generate complex and diverse patterns as a function of a governing parameter and, as such, particularly well suited for severely testing the NN architecture. Indeed, the flexible nature of the used deep learning tools gives the possibility of extending the approach to condition the dynamics and extract other parameters (e.g., remaining in the context of elastically strained alloys, elastic constants in non-homogeneous systems, external loads, plastic contributions, etc.).

An exciting potential application of this framework lies in its use on real experimental images, offering a pathway to bridge the gap between simulation and experiment. By incorporating measurements, this approach could further refine predictions, enabling targeted design and optimization in materials science and semiconductor physics. In fact, the direct, on-the-fly application of ML techniques to experimental data has already begun to emerge \cite{shen_machine-learning-assisted_2024, shen_universal_2024} and is likely to become standard in the coming years.

\section*{Acknowledgements}
FM, RB, and DL acknowledge financial support from ICSC—Centro Nazionale di Ricerca in High-Performance Computing, Big Data and Quantum Computing, funded by the European Union—NextGenerationEU.

\section*{Data availability statement}
The data supporting the findings of this study are available upon reasonable request from the authors. The code used to train the NN model is freely available on GitHub at \url{https://github.com/dlanzo/CRANE}. The dataset used to train and validate the models is available at \url{https://doi.org/10.24435/materialscloud:y7-nj}.

\bibliographystyle{ieeetr} 
\bibliography{references.bib}

\begin{thebibliography}{10}

\bibitem{kok2018anisotropy}
Y.~Kok, X.~P. Tan, P.~Wang, M.~Nai, N.~H. Loh, E.~Liu, and S.~B. Tor,
  ``Anisotropy and heterogeneity of microstructure and mechanical properties in
  metal additive manufacturing: A critical review,'' {\em Materials \& Design},
  vol.~139, pp.~565--586, 2018.

\bibitem{li2020mechanical}
X.~Li, L.~Lu, J.~Li, X.~Zhang, and H.~Gao, ``Mechanical properties and
  deformation mechanisms of gradient nanostructured metals and alloys,'' {\em
  Nature Reviews Materials}, vol.~5, no.~9, pp.~706--723, 2020.

\bibitem{kim2000computation}
Y.-T. Kim, N.~Goldenfeld, and J.~Dantzig, ``Computation of dendritic
  microstructures using a level set method,'' {\em Physical Review E}, vol.~62,
  no.~2, p.~2471, 2000.

\bibitem{elder2007phase}
K.~R. Elder, N.~Provatas, J.~Berry, P.~Stefanovic, and M.~Grant, ``Phase-field
  crystal modeling and classical density functional theory of freezing,'' {\em
  Physical Review B—Condensed Matter and Materials Physics}, vol.~75, no.~6,
  p.~064107, 2007.

\bibitem{moelans2008introduction}
N.~Moelans, B.~Blanpain, and P.~Wollants, ``An introduction to phase-field
  modeling of microstructure evolution,'' {\em Calphad}, vol.~32, no.~2,
  pp.~268--294, 2008.

\bibitem{lyngby_data-driven_2022}
P.~Lyngby and K.~S. Thygesen, ``Data-driven discovery of {2D} materials by deep
  generative models,'' {\em npj Comput Mater}, vol.~8, p.~232, Nov. 2022.

\bibitem{zhao_physics_2023}
Y.~Zhao, E.~M.~D. Siriwardane, Z.~Wu, N.~Fu, M.~Al-Fahdi, M.~Hu, and J.~Hu,
  ``Physics guided deep learning for generative design of crystal materials
  with symmetry constraints,'' {\em npj Comput Mater}, vol.~9, p.~38, Mar.
  2023.

\bibitem{chenebuah_deep_2024}
E.~T. Chenebuah, M.~Nganbe, and A.~B. Tchagang, ``A deep generative modeling
  architecture for designing lattice-constrained perovskite materials,'' {\em
  npj Comput Mater}, vol.~10, p.~198, Aug. 2024.

\bibitem{karpovich_deep_2024}
C.~Karpovich, E.~Pan, and E.~A. Olivetti, ``Deep reinforcement learning for
  inverse inorganic materials design,'' {\em npj Comput Mater}, vol.~10,
  p.~287, Dec. 2024.

\bibitem{ward_general-purpose_2016}
L.~Ward, A.~Agrawal, A.~Choudhary, and C.~Wolverton, ``A general-purpose
  machine learning framework for predicting properties of inorganic
  materials,'' {\em npj Comput Mater}, vol.~2, p.~16028, Aug. 2016.

\bibitem{dunn_benchmarking_2020}
A.~Dunn, Q.~Wang, A.~Ganose, D.~Dopp, and A.~Jain, ``Benchmarking materials
  property prediction methods: the {Matbench} test set and {Automatminer}
  reference algorithm,'' {\em npj Comput Mater}, vol.~6, p.~138, Sept. 2020.

\bibitem{de_breuck_materials_2021}
P.-P. De~Breuck, G.~Hautier, and G.-M. Rignanese, ``Materials property
  prediction for limited datasets enabled by feature selection and joint
  learning with {MODNet},'' {\em npj Comput Mater}, vol.~7, p.~83, June 2021.

\bibitem{bartok_gaussian_2010}
A.~P. Bartók, M.~C. Payne, R.~Kondor, and G.~Csányi, ``Gaussian
  {Approximation} {Potentials}: {The} {Accuracy} of {Quantum} {Mechanics},
  without the {Electrons},'' {\em Phys. Rev. Lett.}, vol.~104, p.~136403, Apr.
  2010.

\bibitem{behler_atom-centered_2011}
J.~Behler, ``Atom-centered symmetry functions for constructing high-dimensional
  neural network potentials,'' {\em The Journal of Chemical Physics}, vol.~134,
  p.~074106, Feb. 2011.

\bibitem{fantasia_development_2024}
A.~Fantasia, F.~Rovaris, O.~Abou El~Kheir, A.~Marzegalli, D.~Lanzoni,
  L.~Pessina, P.~Xiao, C.~Zhou, L.~Li, G.~Henkelman, E.~Scalise, and
  F.~Montalenti, ``Development of a machine learning interatomic potential for
  exploring pressure-dependent kinetics of phase transitions in germanium,''
  {\em The Journal of Chemical Physics}, vol.~161, p.~014110, July 2024.

\bibitem{shen_-situ_2024}
C.~Shen, W.~Zhan, S.~Pan, H.~Hao, N.~Zhuo, K.~Xin, H.~Cong, C.~Xu, B.~Xu, T.~K.
  Ng, S.~Chen, C.~Xue, F.~Liu, Z.~Wang, and C.~Zhao, ``In-situ
  {Self}-optimization of {Quantum} {Dot} {Emission} for {Lasers} by
  {Machine}-{Learning} {Assisted} {Epitaxy},'' Nov. 2024.

\bibitem{shen_autonomous_2024}
C.~Shen, W.~Zhan, H.~Sun, K.~Xin, B.~Xu, Z.~Wang, and C.~Zhao, ``Autonomous,
  {Self}-driving {Multi}-{Step} {Growth} of {Semiconductor} {Heterostructures}
  {Guided} by {Machine} {Learning},'' Aug. 2024.

\bibitem{strayerADDMANL2022}
S.~T. Strayer, W.~J.~F. Templeton, F.~X. Dugast, S.~P. Narra, and A.~C. To,
  ``Accelerating {{High-Fidelity Thermal Process Simulation}} of {{Laser Powder
  Bed Fusion}} via the {{Computational Fluid Dynamics Imposed Finite Element
  Method}} ({{CIFEM}}),'' {\em Additive Manufacturing Letters}, vol.~3,
  p.~100081, 2022.

\bibitem{lanzoniAPLML2024}
D.~Lanzoni, F.~Rovaris, L.~{Mart{\'i}n-Encinar}, A.~Fantasia, R.~Bergamaschini,
  and F.~Montalenti, ``Accelerating simulations of strained-film growth by deep
  learning: {{Finite}} element method accuracy over long time scales,'' {\em
  APL Machine Learning}, vol.~2, p.~036108, 2024.

\bibitem{montes_de_oca_zapiain_accelerating_2021}
D.~Montes De Oca~Zapiain, J.~A. Stewart, and R.~Dingreville, ``Accelerating
  phase-field-based microstructure evolution predictions via surrogate models
  trained by machine learning methods,'' {\em npj Comput Mater}, vol.~7, p.~3,
  Jan. 2021.

\bibitem{hu_accelerating_2022}
C.~Hu, S.~Martin, and R.~Dingreville, ``Accelerating phase-field predictions
  via recurrent neural networks learning the microstructure evolution in latent
  space,'' {\em Computer Methods in Applied Mechanics and Engineering},
  vol.~397, p.~115128, July 2022.

\bibitem{wu_emulating_2023}
P.~Wu, A.~S. Iquebal, and K.~Ankit, ``Emulating microstructural evolution
  during spinodal decomposition using a tensor decomposed convolutional and
  recurrent neural network,'' {\em Computational Materials Science}, vol.~224,
  p.~112187, May 2023.

\bibitem{ahmad_integrated_2024}
O.~Ahmad, R.~Maurya, R.~Mukherjee, and S.~Bhowmick, ``Integrated {Phase}
  {Field} and {Machine} {Learning} {Study} of {Microstructure} {Evolution}
  during {Interface}-{Controlled} {Spinodal} {Decomposition},'' {\em SSP},
  vol.~357, pp.~101--106, June 2024.

\bibitem{lanzoni_extreme_2024}
D.~Lanzoni, A.~Fantasia, R.~Bergamaschini, O.~Pierre-Louis, and F.~Montalenti,
  ``Extreme time extrapolation capabilities and thermodynamic consistency of
  physics-inspired neural networks for the {3D} microstructure evolution of
  materials via {Cahn}–{Hilliard} flow,'' {\em Mach. Learn.: Sci. Technol.},
  vol.~5, p.~045017, Dec. 2024.

\bibitem{ren_numerical_2022}
L.~Ren, S.~Geng, P.~Jiang, S.~Gao, and C.~Han, ``Numerical simulation of
  dendritic growth during solidification process using multiphase-field model
  aided with machine learning method,'' {\em Calphad}, vol.~78, p.~102450,
  Sept. 2022.

\bibitem{ren_phase-field_2022}
Y.~Ren, K.~Zhang, Y.~Zhou, and Y.~Cao, ``Phase-{Field} {Simulation} and
  {Machine} {Learning} {Study} of the {Effects} of {Elastic} and {Plastic}
  {Properties} of {Electrodes} and {Solid} {Polymer} {Electrolytes} on the
  {Suppression} of {Li} {Dendrite} {Growth},'' {\em ACS Appl. Mater.
  Interfaces}, vol.~14, pp.~30658--30671, July 2022.

\bibitem{lee_recent_2023}
H.~Lee and D.~Kim, ``Recent {Computational} {Approaches} for {Accelerating}
  {Dendrite} {Growth} {Prediction}: {A} {Short} {Review},'' {\em Multiscale
  Sci. Eng.}, vol.~5, pp.~119--125, Dec. 2023.

\bibitem{wang_modeling_2024}
X.~Wang, S.~Li, and F.~Liu, ``Modeling for free dendrite growth based on
  physically-informed machine learning method,'' {\em Scripta Materialia},
  vol.~242, p.~115918, Mar. 2024.

\bibitem{choi_accelerating_2024}
J.~Y. Choi, T.~Xue, S.~Liao, and J.~Cao, ``Accelerating phase-field simulation
  of three-dimensional microstructure evolution in laser powder bed fusion with
  composable machine learning predictions,'' {\em Additive Manufacturing},
  vol.~79, p.~103938, Jan. 2024.

\bibitem{alhadaNPJCM2024}
K.~{Alhada--Lahbabi}, D.~Deleruyelle, and B.~Gautier, ``Machine learning
  surrogate for {{3D}} phase-field modeling of ferroelectric tip-induced
  electrical switching,'' {\em npj Computational Materials}, vol.~10, p.~197,
  2024.

\bibitem{yang_self-supervised_2021}
K.~Yang, Y.~Cao, Y.~Zhang, S.~Fan, M.~Tang, D.~Aberg, B.~Sadigh, and F.~Zhou,
  ``Self-supervised learning and prediction of microstructure evolution with
  convolutional recurrent neural networks,'' {\em Patterns}, vol.~2, p.~100243,
  May 2021.

\bibitem{fan2024accelerate}
S.~Fan, A.~L. Hitt, M.~Tang, B.~Sadigh, and F.~Zhou, ``Accelerate
  microstructure evolution simulation using graph neural networks with adaptive
  spatiotemporal resolution,'' {\em Machine Learning: Science and Technology},
  vol.~5, no.~2, p.~025027, 2024.

\bibitem{guptaARXIV2022}
J.~K. Gupta and J.~Brandstetter, ``Towards {{Multi-spatiotemporal-scale
  Generalized PDE Modeling}},'' 2022.

\bibitem{oommen_rethinking_2023}
V.~Oommen, K.~Shukla, S.~Desai, R.~Dingreville, and G.~E. Karniadakis,
  ``Rethinking materials simulations: {Blending} direct numerical simulations
  with neural operators,'' 2023.
\newblock Version Number: 1.

\bibitem{lanzoni_morphological_2022}
D.~Lanzoni, M.~Albani, R.~Bergamaschini, and F.~Montalenti, ``Morphological
  evolution via surface diffusion learned by convolutional, recurrent neural
  networks: {Extrapolation} and prediction uncertainty,'' {\em Phys. Rev.
  Materials}, vol.~6, p.~103801, Oct. 2022.

\bibitem{cahnACTAMETAL1961}
J.~W. Cahn, ``On spinodal decomposition,'' {\em Acta Metallurgica}, vol.~9,
  no.~9, pp.~795--801, 1961.

\bibitem{langer_theory_1973}
J.~Langer and M.~Bar-on, ``Theory of early-stage spinodal decomposition,'' {\em
  Annals of Physics}, vol.~78, pp.~421--452, June 1973.

\bibitem{chenBOOK1994}
L.~Q. Chen, Y.~Z. Wang, and A.~G. Khachaturyan, ``Morphology
  {{Transformations}} in {{Ordering}} and {{Phase Separating Materials}},'' in
  {\em Statics and {{Dynamics}} of {{Alloy Phase Transformations}}} (P.~E.~A.
  Turchi and A.~Gonis, eds.), pp.~587--604, Boston, MA: Springer US, 1994.

\bibitem{fratzl_modeling_1999}
P.~Fratzl, O.~Penrose, and J.~L. Lebowitz, ``Modeling of {Phase} {Separation}
  in {Alloys} with {Coherent} {Elastic} {Misfit},'' {\em Journal of Statistical
  Physics}, vol.~95, no.~5/6, pp.~1429--1503, 1999.

\bibitem{kwon_coarsening_2007}
Y.~Kwon, K.~Thornton, and P.~W. Voorhees, ``Coarsening of bicontinuous
  structures via nonconserved and conserved dynamics,'' {\em Phys. Rev. E},
  vol.~75, p.~021120, Feb. 2007.

\bibitem{andrews_effect_2020}
W.~B. Andrews, K.~L.~M. Elder, P.~W. Voorhees, and K.~Thornton, ``Effect of
  transport mechanism on the coarsening of bicontinuous structures: {A}
  comparison between bulk and surface diffusion,'' {\em Phys. Rev. Materials},
  vol.~4, p.~103401, Oct. 2020.

\bibitem{rundman_early_1967}
K.~Rundman and J.~Hilliard, ``Early stages of spinodal decomposition in an
  aluminum-zinc alloy,'' {\em Acta Metallurgica}, vol.~15, pp.~1025--1033, June
  1967.

\bibitem{xu_stabilizing_2024}
W.~Xu, Y.~Zhong, X.~Li, and K.~Lu, ``Stabilizing {Supersaturation} with
  {Extreme} {Grain} {Refinement} in {Spinodal} {Aluminum} {Alloys},'' {\em
  Advanced Materials}, vol.~36, p.~2303650, Apr. 2024.

\bibitem{collins_spinodal_2020}
D.~M. Collins, N.~D’Souza, C.~Panwisawas, C.~Papadaki, G.~D. West, A.~Kostka,
  and P.~Kontis, ``Spinodal decomposition versus classical $\gamma'$ nucleation
  in a nickel-base superalloy powder: {An} in-situ neutron diffraction and
  atomic-scale analysis,'' {\em Acta Materialia}, vol.~200, pp.~959--970, Nov.
  2020.

\bibitem{kim_spinodal_2002}
H.~Kim and P.~C. McIntyre, ``Spinodal decomposition in amorphous
  metal–silicate thin films: {Phase} diagram analysis and interface effects
  on kinetics,'' {\em Journal of Applied Physics}, vol.~92, pp.~5094--5102,
  Nov. 2002.

\bibitem{ban_spinodal_2023}
M.~Ban, D.~Woo, J.~Hwang, S.~Kim, and J.~Lee, ``Spinodal
  {Decomposition}-{Driven} {Structural} {Hierarchy} of {Mesoporous} {Inorganic}
  {Materials} for {Energy} {Applications},'' {\em Accounts of Chemical
  Research}, vol.~56, pp.~3428--3440, Dec. 2023.

\bibitem{kumar_inverse-designed_2020}
S.~Kumar, S.~Tan, L.~Zheng, and D.~M. Kochmann, ``Inverse-designed spinodoid
  metamaterials,'' {\em npj Computational Materials}, vol.~6, p.~73, June 2020.

\bibitem{zhengCMAME2021}
L.~Zheng, S.~Kumar, and D.~M. Kochmann, ``Data-driven topology optimization of
  spinodoid metamaterials with seamlessly tunable anisotropy,'' {\em Computer
  Methods in Applied Mechanics and Engineering}, vol.~383, p.~113894, 2021.

\bibitem{nishimoriPRB1990}
H.~Nishimori and A.~Onuki, ``Pattern formation in phase-separating alloys with
  cubic symmetry,'' {\em Physical Review B}, vol.~42, no.~1, pp.~980--983,
  1990.

\bibitem{zhuMSME2001}
J.~Zhu, L.-Q. Chen, and J.~Shen, ``Morphological evolution during phase
  separation and coarsening with strong inhomogeneous elasticity,'' {\em
  Modelling and Simulation in Materials Science and Engineering}, vol.~9,
  no.~6, p.~499, 2001.

\bibitem{cahnACTAMETAL1962}
J.~W. Cahn, ``On spinodal decomposition in cubic crystals,'' {\em Acta
  Metallurgica}, vol.~10, no.~3, pp.~179--183, 1962.

\bibitem{provatas_phasefield_2010}
N.~Provatas and K.~Elder, {\em Phase‐{Field} {Methods} in {Materials}
  {Science} and {Engineering}}.
\newblock Wiley, 1~ed., Oct. 2010.

\bibitem{garcke2003}
H.~Garcke, S.~{Maier-Paape}, and U.~Weikard, ``Spinodal {{Decomposition}} in
  the {{Presence}} of {{Elastic Interactions}},'' in {\em Geometric
  {{Analysis}} and {{Nonlinear Partial Differential Equations}}}
  (S.~Hildebrandt and H.~Karcher, eds.), pp.~603--635, Berlin, Heidelberg:
  Springer, 2003.

\bibitem{mura1987}
T.~Mura, {\em Micromechanics of Defects in Solids}, vol.~3 of {\em Mechanics of
  {{Elastic}} and {{Inelastic Solids}}}.
\newblock Dordrecht: Springer Netherlands, 1987.

\bibitem{khachaturyan2008theory}
A.~Khachaturyan, {\em Theory of Structural Transformations in Solids}.
\newblock Dover Books on Engineering Series, Dover Publications, 2008.

\bibitem{cahn_free_1958}
J.~W. Cahn and J.~E. Hilliard, ``Free {Energy} of a {Nonuniform} {System}. {I}.
  {Interfacial} {Free} {Energy},'' {\em The Journal of Chemical Physics},
  vol.~28, pp.~258--267, Feb. 1958.

\bibitem{cahn_phase_1965}
J.~W. Cahn, ``Phase {Separation} by {Spinodal} {Decomposition} in {Isotropic}
  {Systems},'' {\em The Journal of Chemical Physics}, vol.~42, pp.~93--99, Jan.
  1965.

\bibitem{garcke2006}
H.~Garcke, M.~Lenz, B.~Niethammer, M.~Rumpf, and U.~Weikard, ``Multiple
  {{Scales}} in {{Phase Separating Systems}} with {{Elastic Misfit}},'' in {\em
  Analysis, {{Modeling}} and {{Simulation}} of {{Multiscale Problems}}}
  (A.~Mielke, ed.), (Berlin, Heidelberg), pp.~153--178, Springer, 2006.

\bibitem{wang_kinetics_1993}
Y.~Wang, L.-Q. Chen, and A.~Khachaturyan, ``Kinetics of strain-induced
  morphological transformation in cubic alloys with a miscibility gap,'' {\em
  Acta Metallurgica et Materialia}, vol.~41, pp.~279--296, Jan. 1993.

\bibitem{cahn_cubicACTAMETAL1962}
J.~W. Cahn, ``On spinodal decomposition in cubic crystals,'' {\em Acta
  Metallurgica}, vol.~10, no.~3, pp.~179--183, 1962.

\bibitem{vos_relationship_1966}
K.~K.~D. Vos, ``The relationship between microstructure and magnetic properties
  of alnico alloys,'' 1966.

\bibitem{ardell_modulated_1966}
A.~Ardell and R.~Nicholson, ``On the modulated structure of aged {Ni}-{Al}
  alloys,'' {\em Acta Metallurgica}, vol.~14, pp.~1295--1309, Oct. 1966.

\bibitem{perlin_image_1985}
K.~Perlin, ``An image synthesizer,'' {\em SIGGRAPH Comput. Graph.}, vol.~19,
  pp.~287--296, July 1985.

\bibitem{chung_empirical_2014}
J.~Chung, C.~Gulcehre, K.~Cho, and Y.~Bengio, ``Empirical {Evaluation} of
  {Gated} {Recurrent} {Neural} {Networks} on {Sequence} {Modeling},'' 2014.

\bibitem{circular_padding}
Schubert S, Neubert P, Pöschmann J and Protzel P 2019 2019 IEEE Intelligent
  Vehicles Symposium (IV) (IEEE) pp 653–60.

\bibitem{kingma_adam_2014}
D.~P. Kingma and J.~Ba, ``Adam: {A} {Method} for {Stochastic} {Optimization},''
  2014.

\bibitem{salvalaglioCGD2015}
M.~Salvalaglio, R.~Backofen, R.~Bergamaschini, F.~Montalenti, and A.~Voigt,
  ``Faceting of {{Equilibrium}} and {{Metastable Nanostructures}}: {{A
  Phase-Field Model}} of {{Surface Diffusion Tackling Realistic Shapes}},''
  {\em Crystal Growth \& Design}, vol.~15, no.~6, pp.~2787--2794, 2015.

\bibitem{bengio_curriculum_2009}
Y.~Bengio, J.~Louradour, R.~Collobert, and J.~Weston, ``Curriculum learning,''
  in {\em Proceedings of the 26th {Annual} {International} {Conference} on
  {Machine} {Learning}}, (Montreal Quebec Canada), pp.~41--48, ACM, June 2009.

\bibitem{shen_machine-learning-assisted_2024}
C.~Shen, W.~Zhan, K.~Xin, M.~Li, Z.~Sun, H.~Cong, C.~Xu, J.~Tang, Z.~Wu, B.~Xu,
  Z.~Wei, C.~Xue, C.~Zhao, and Z.~Wang, ``Machine-learning-assisted and
  real-time-feedback-controlled growth of {InAs}/{GaAs} quantum dots,'' {\em
  Nat Commun}, vol.~15, p.~2724, Mar. 2024.

\bibitem{shen_universal_2024}
C.~Shen, W.~Zhan, J.~Tang, Z.~Wu, B.~Xu, C.~Zhao, and Z.~Wang, ``Universal
  {Deoxidation} of {Semiconductor} {Substrates} {Assisted} by {Machine}
  {Learning} and {Real}-{Time} {Feedback} {Control},'' {\em ACS Appl. Mater.
  Interfaces}, vol.~16, pp.~18213--18221, Apr. 2024.

\end{thebibliography}

\end{document}